\documentstyle[floats,twocolumn,epsf,prl,aps]{revtex}

\def\figun{%
\begin{figure}
\[\epsfbox{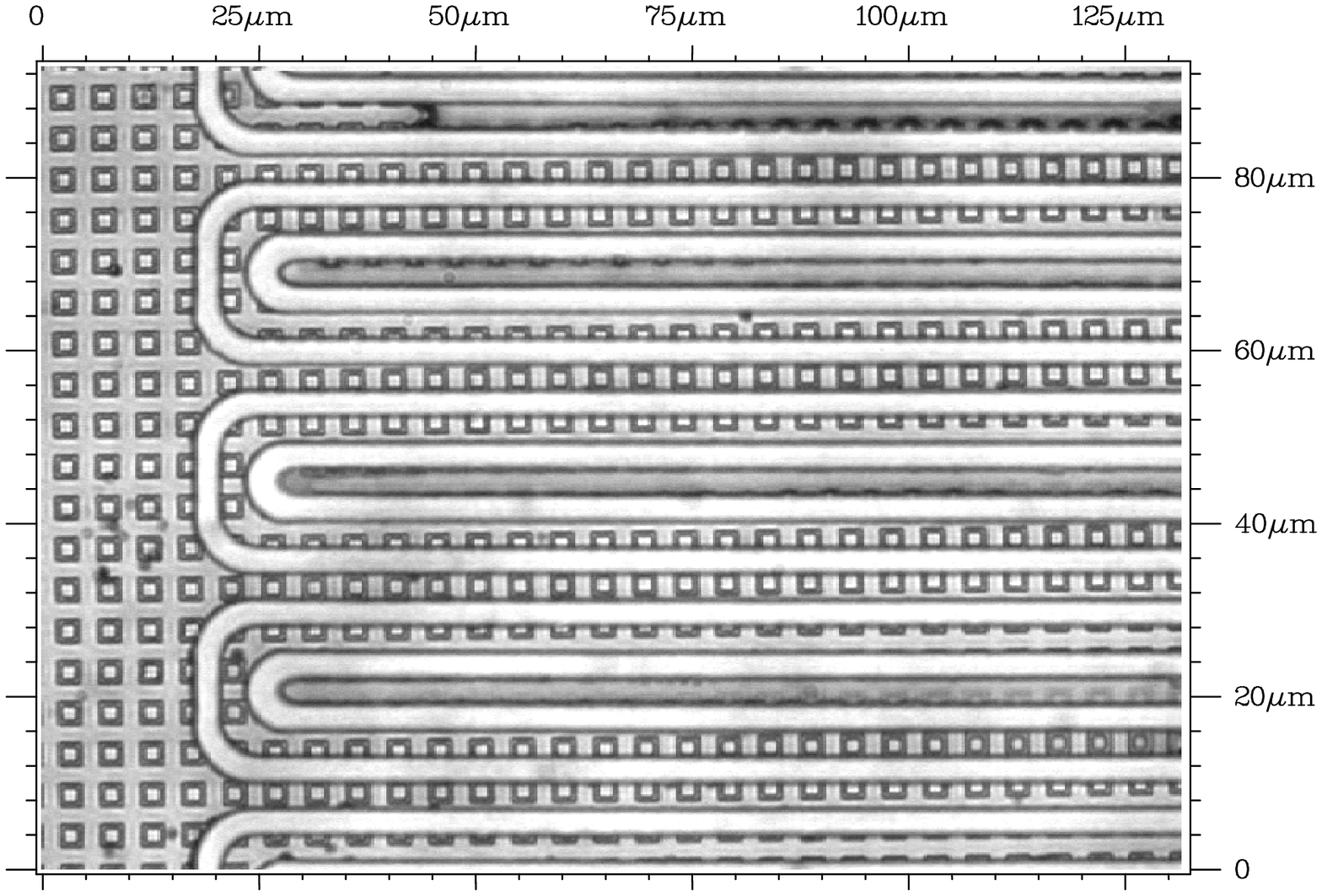}\]%
\caption{Optical photography of a piece of the resonator coupled to the
$GaAs/GaAlAs$ mesoscopic rings. One
notes the two folded Nb lines ($1\mu$m thick, $2\mu$m wide and $20$cm long) on
the sapphire substrate.
\label{fig_PhotoLi}}
\end{figure}
}

\def\figdeux{%
\begin{figure}
\[\epsfbox{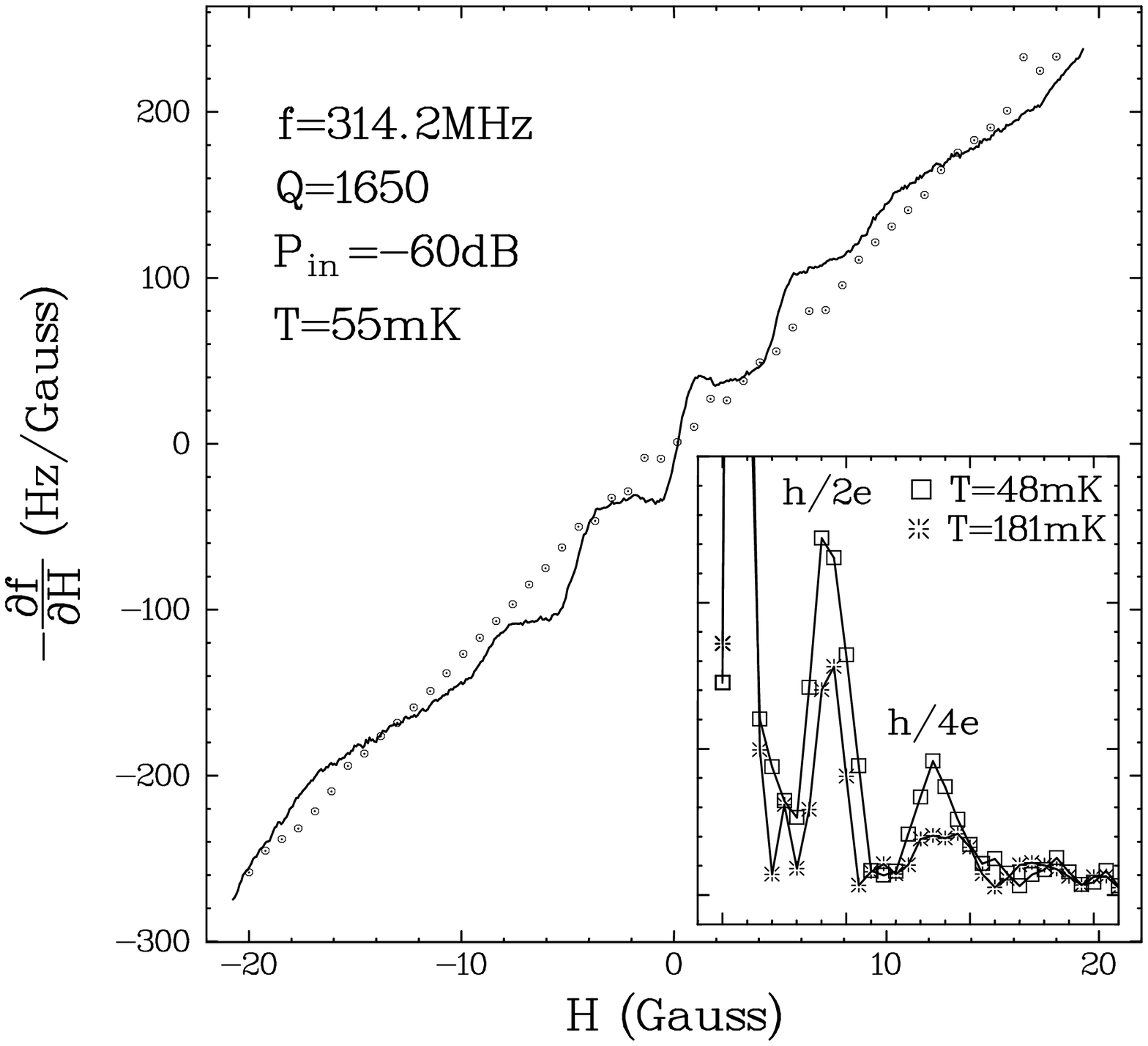}\]%
\caption{Evolution of the derivative of the fundamental resonance
frequency $f_1$ of the line as a function of the dc magnetic
field. The linear background observed in the absence of the rings (open
symbols)  corresponds to the diamagnetism of the Nb, on
which are superimposed the $h/2e$ oscillations due to the mesoscopic rings.
This curve is averaged $40$ times. Inset: Fourier transform of the signal in
the presence of the rings for two different
temperatures.
\label{fig_F0Phi}}
\end{figure}
}

\def\figtrois{%
\begin{figure}
\[\epsfbox{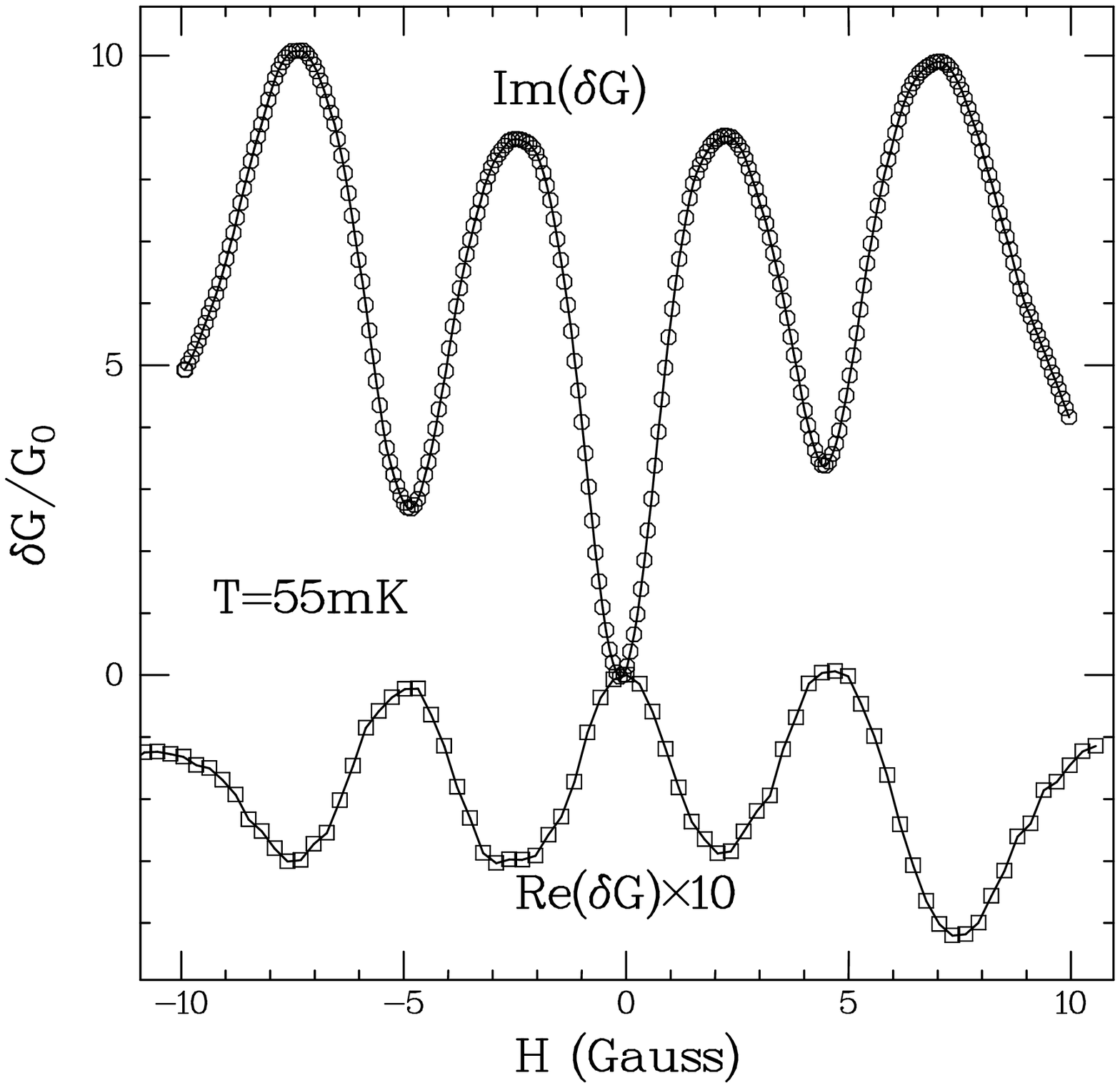}\]%
\caption{Magnetic field dependence of the imaginary
and real components of the conductance of the rings expressed in
units of the Drude conductance $G_0$. These two curves have been
obtained by integrating the measured quantities $\partial f_1/\partial
H$ and $\partial Q_1/\partial H$ after substraction of the
background.
\label{fig_fQPhi}}
\end{figure}
}

\def\figquatre{%
\begin{figure}
\[\epsfbox{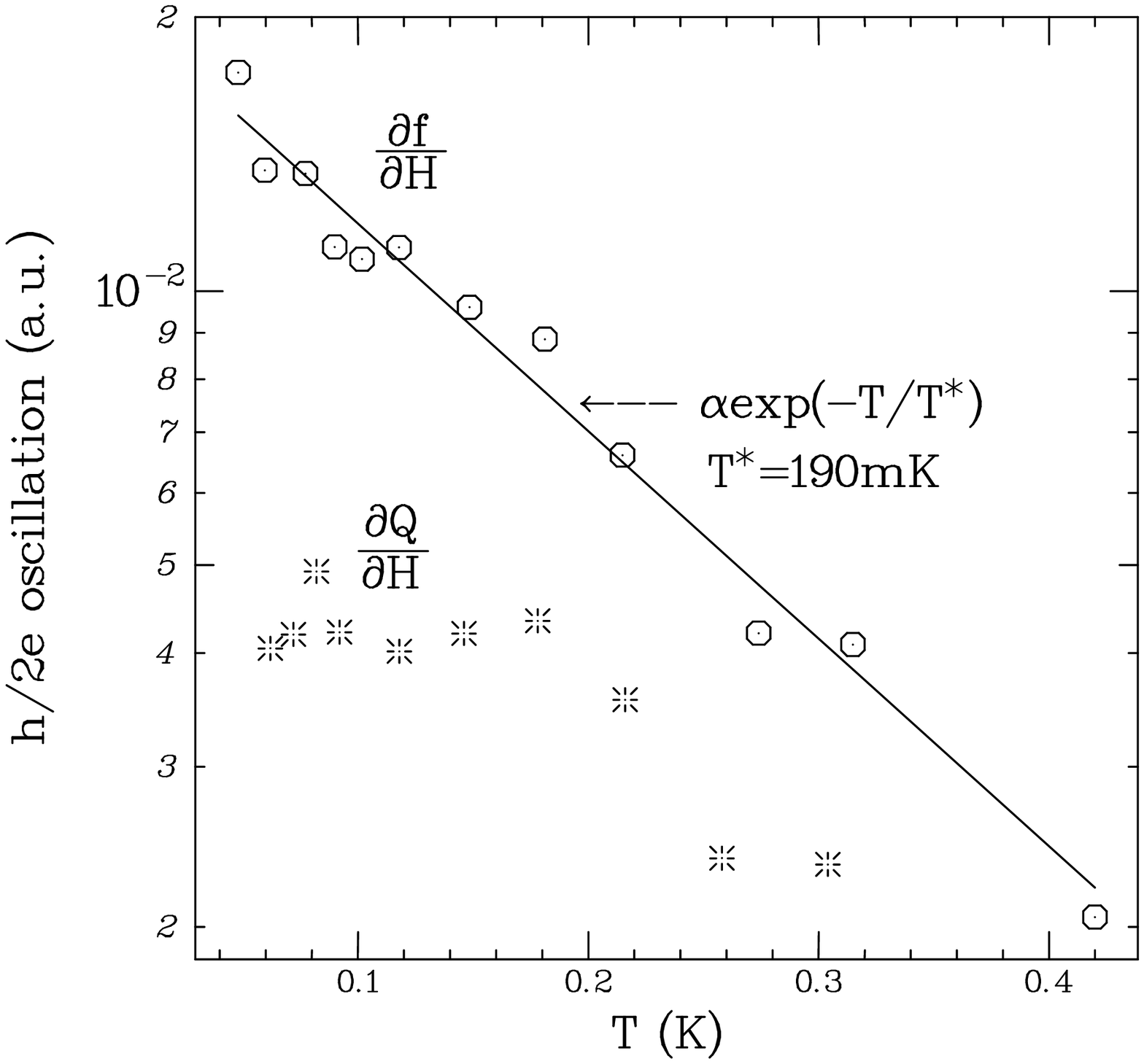}\]%
\caption{ Temperature dependence of the $h/2e$ periodic components of
$\partial f_1/\partial H$ and $\partial Q_1/\partial H$.
\label{fig_fQT}}
\end{figure}
}

\title{Dynamic response of isolated Aharonov-Bohm rings coupled to an
electromagnetic resonator}

\author{Bertrand Reulet$^1$, Michel Ramin $^1$, H\'el\`ene Bouchiat$^1$ and
Dominique Mailly$^{2}$}

\address{$^1$Laboratoire de Physique des Solides, Associ\'e au CNRS, B\^at 510,
Universit\'e  Paris--Sud, 91405, Orsay, France.\\ $^2$C.N.R.S Laboratoire de
Microstructures et de Micro\'electronique, 196, Avenue Ravera, 92220, Bagneux,
France\\%
\parbox{14cm}{\medskip\rm\small%
We have measured the flux dependence of both real and imaginary conductance of
$GaAs/GaAlAs$ isolated mesoscopic rings at 310 MHz. The rings are coupled to a
highly
sensitive electromagnetic superconducting micro-resonator and lead to a
perturbation of the resonance frequency and quality factor. This experiment
provides a new tool for the investigation of the conductance of mesoscopic
systems without any connection to invasive probes. It can be compared with
recent theoretical predictions emphasizing the differences between isolated and
connected geometries and the relation  between ac conductance and persistent
currents. We observe $\Phi_0/2$ periodic oscillations on both  components of
the magnetoconductance. The oscillations of the imaginary conductance whose
sign
corresponds to diamagnetism in zero field, are  3 times larger than the
Drude conductance $G_0$. The real part of the periodic magnetoconductance is
of the  order of $0.2 G_0$ and is apparently negative in low field. It is
thus notably different from the weak localisation oscillations observed in
connected rings, which  are  much smaller and  opposite in sign.
}%
}

\begin{document}

\maketitle


	Mesoscopic metallic rings present a spectacular thermodynamic property~:
they carry a persistent non dissipative current when they are threaded by a
magnetic flux \cite{art_LDDB,art_ManipWebb,art_ManipBenoit}. The
existence of such a persistent current is a consequence of the coherence of the
electronic wave functions along the ring. However unlike a superconductor, when
connected to a voltage source, the same  rings  present a finite ohmic
conductance whose value is close to its classical expectation given by the
Drude formula, which depends only on the elastic scattering time (quantum
interference effects give rise to contributions which are only a
small fraction of this main classical contribution in the metallic diffusive
regime). It has already been pointed out a number of times\cite{lan},
that the existence
of a finite ohmic resistance for a phase coherent sample is not paradoxical
when
one properly takes into account the influence of the measuring leads. These
macroscopic wires connected to the sample play indeed the role of incoherent
reservoirs of electrons where thermalisation of the electrons and dissipation
take place. Such a strong coupling with a reservoir of electrons can be avoided
by studying the current response of a mesoscopic ring to a time dependant flux,
which induces an electric field along the ring. Since the early work of
B\"uttiker et al. \cite{art_ButImrLan,art_LanBut_PRL85,art_But_AnnNYAS}
subsequently generalized by a number
of authors\cite{art_ImrShi,art_TB,art_RB,art_KRBG,art_KG}, it has
been shown that the conductance measured by this last technique on an isolated
ring is indeed fundamentally different from the conductance of the same sample
connected to a voltage source. It essentially depends on the inelastic time
$\tau_{in}$ (which describes the coupling of the electrons to the thermal
bath). Furthermore it is strongly related to the presence of persistent
currents
through the ring.

 	In its ac version the experiment consists in measuring the complex magnetic
susceptibility of the rings $\chi(\omega)=\chi'(\omega)+ i\chi''(\omega)$
submitted to a small ac flux superimposed on a dc one $\Phi$.
In the linear response limit, this susceptibility is related to the complex
ac conductance of the rings $G$ by
$\chi(\omega)\propto i\omega G(\omega)$. Let us summarize the main theoretical
predictions \cite {art_RB,art_KRBG,art_KG}.
An important energy scale for the dynamics of the system is its thermalisation
time $\gamma^{-1}\approx\tau_{in}$. In the adiabatic limit $\omega<\gamma$,
$\Im m(G)$ reduces to the derivative  of the persistent current, while a
relaxation term occurs at higher frequency. In the continuous
spectrum limit $\gamma\gg\Delta$ (where $\Delta$ denotes the mean level
spacing) and zero frequency, $\Re e(G)$ is given by
$\Re e(G)=G_0+\delta G(\Phi)$ where
$\delta G\approx G_0\frac\Delta\gamma\ll G_0$ is the $\Phi_0/2$ periodic
Altshuler Aronov and Spivak (AAS) weak localisation correction, positive in
weak field\cite{art_AAS,art_ShaSha}. Nevertheless the same quantity in the
discrete
spectrum limit $\gamma\ll\Delta$ (and in the canonical statistical ensemble
corresponding to isolated objects) may present oscillations of opposite
sign (for $T<\Delta$, $\omega<\gamma$) and amplitude of the order of $G_0$.
These oscillations are predicted to reverse sign and become of the order
of $\frac\Delta\gamma G_0$ when the temperature $T$ increases.

Motivated by these findings, we have designed an experiment to measure the
complex ac conductance of an array of $GaAs/GaAlAs$  isolated rings. The
discrete
spectrum is much easier to reach in these samples, where $\Delta$ is of the
order of a few tens of mK, than in metallic ones of comparable sizes,
corresponding to
$\Delta$ in the  microkelvin range. The sample is an array of $10^5$ isolated
square rings $2\mu$m on a side, made  using e-beam lithography. The electronic
parameters of the rings are obtained from transport measurements done on
connected rings and wires made using the same process as the isolated rings.
Moreover, because of depletion effects, the real width of the wires etched in
the 2D electron gas is substantially smaller than the nominal one, and must be
determined by weak localisation measurements \cite{art_RBM_epl}. From these
measurements we deduced the following parameters~: $\Delta=35{\rm mK}\quad
E_c=hD/L^2=200{\rm mK}\quad M=17 \quad l_{tr}=3\mu{\rm m}\quad M_{eff}=4 \quad
L_\phi(T=50{\rm mK})=7\mu{\rm m} $ where $M$ denotes the number of channels of
the rings, $M_{eff}$ their effective number, $l_{tr}$ the transport
length and $L_\phi(T)$ the (temperature dependant) phase coherence length of
electrons. The electronic motion is then diffusive along the rings and
ballistic in the transverse direction. In terms of frequency, the energies
are:$\Delta=630{\rm MHz}\quad E_c=4.2{\rm GHz}$

 This determines the interesting range of frequency: from a few hundreds of
megahertz ($\omega<\Delta$) to a few gigahertz ($\omega\approx E_c$). The
inelastic parameter $\gamma$, of course, cannot be a priori deduced from such
transport experiments, since it represents a property of the isolated rings.
Nevertheless, assuming than $\gamma$ is of the order of $\hbar/\tau_\phi$
(where $\tau_\phi$ is the phase coherence time of electrons measured by weak
localisation in wires having the same width than the rings), we expect that
it is smaller than $\Delta$ below $50$mK (such an assumption is in agreement
with the results obtained by Sivan et al. \cite{sivan} on the tunnel
spectroscopy measurements of quantum dots).

Our aim was to be able to detect the in-phase and out-of-phase response of this
array of rings to a small magnetic excitation. We recall that such an
experiment which deals with electrically isolated objects is very different
from the ac measurement of the complex
conductance of connected rings \cite{art_Pieper}. Since the estimated amplitude
of the signal was extremely small, we had to design a special experimental
setup. We have used a resonant technique in which the rings are magnetically
coupled to an electromagnetic multi-mode resonator, whose performances are
affected by the perturbations due to the rings. The resonance frequencies
$f_n$ and  quality factors $Q_n$   are modified by the presence of the rings
according to:

\begin{equation}
\left\{
\begin{array}{rcl}
\displaystyle\frac{\delta f_n}{f_n}&=&\displaystyle
\frac{2\pi N{\cal M}^2f_n\Im m[G(f_n)]}{\cal L} \\
\displaystyle\frac{\delta Q_n}{Q_n^2}&=&
\displaystyle\frac{2\pi N{\cal M}^2f_n \Re e[G(f_n)]}{\cal L}
\end{array}\right.
\label{eqM}
\end{equation}
where $N$ is the number of rings and $\cal M$ the mutual inductance between
one ring and the resonator of self-inductance $\cal L$.
Considerations of coupling optimization between the samples and the detector,
have lead us to use the meander strip line resonator depicted on fig.
\ref{fig_PhotoLi} on the top of which the array of rings is deposited. In this
geometry each ring is close to the resonating line, which ensures a good mutual
coupling between them. The line, open at both ends, has resonances each time
its length is a multiple value of $\lambda_0/2$, where $\lambda_0$ is the
electromagnetic wavelength.

\figun

Typical superconducting Nb resonators produced on sapphire substrates  have a
fundamental resonant
frequency of $380$Mhz and a $Q=80,000$  at temperatures below $1$K. The
sensitivity of our experiment is determined by the precision with which we can
detect a small deviation of $f_n$ and $Q_n$, actually~:
$\delta f_n/f_n\approx10^{-9}$ and $\delta Q_n/Q_n^2\approx10^{-10}$,
for an injected power of 1nW, which avoids the heating of electrons
and corresponds to an ac magnetic flux less than $0.1\Phi_0$
through the rings (more technical details will be published somewhere else
\cite{trieste}).

Ideally all rings should be exposed to the same ac magnetic field and therefore
should have a very well controlled position tightly coupled to the resonator.
But for the moment this is very difficult to achieve, since for reasons of
lithography the line and the rings are sitting on different substrates.
However, as long as we are concerned only with the linear response, it is not
required that all the rings experience the same ac field (as soon as it is
small enough). The problem of the homogeneity of the dc magnetic field is
somewhat more serious. Since the characteristic signature of the effects we are
looking for, are periodic oscillations with the dc flux through the rings, it
is crucial that they see essentially the same dc field. Due to the Meissner
effect, the dc field just above the resonator is strongly inhomogeneous
spatially. These field inhomogeneities decrease however exponentially with the
distance between the rings and line substrates and are reduced to $10$\% when
a  $1.5\mu$m thick, mylar film is inserted between the detector and the rings
substrate. We have estimated in
this geometry the typical mutual inductance $\cal M$ between one ring and the
resonator
to be of the order of $1.5\;10^{-13}$H.
We have checked this value by measuring the susceptibility of an array of
superconducting aluminum rings. The most serious difficulty we had to overcome
in order
to realize this experiment is the existence of spurious losses coming from the
partially etched $GaAlAs$ top layer of the heterostructure. The first  attempt
was done with very slightly etched samples, we observed a drop of
the quality factor of the resonator from $80,000$ to $10$. By etching the
samples more deeply we could decrease these losses by a factor $100$ and we
obtained the results depicted below where $Q=1650$ for the fundamental
frequency. It was not possible yet to work on the higher harmonics . We hope to
reduce further these residual losses in future experiments.

\figdeux
Let us now describe the magnetic field and temperature dependence of
the complex susceptibility of the array of $GaAs/GaAlAs$ rings
measured at $310$MHz. The measurements were done using two different
resonators and were reproducible from one resonator to the next. The
dc magnetic field was modulated at $3$Hz with an amplitude of
$1$Gauss. The resulting signals are proportional to the derivatives
of $f_1$ and $Q_1$ with respect to the dc magnetic field. In fig.
\ref{fig_F0Phi} we show the field dependence of
$-\frac{\partial f_1}{\partial H}$ averaged 40 times.
One clearly notes in low field the
oscillations associated with the rings superimposed on the linear
dependence corresponding to the diamagnetism of the niobium. The 5
Gauss periodicity corresponds to a flux of amplitude $\Phi_0/2=h/2e$
in the squares. The oscillations are not visible at fields larger
than 10 Gauss which this is due to the rather small aspect ratio of
the rings, $1\Phi_0$ through the area of the wires corresponds indeed
to 15 Gauss. One deduces from fig.\ref{fig_F0Phi}~:

\begin{equation}
-\frac{\partial f_1}{\partial H}=\alpha
H+\beta_1\sin4\pi\frac\Phi{\Phi_0}+\beta_2\sin8\pi\frac\Phi{\Phi_0}
\end{equation}
with $\alpha=13{\rm Hz/Gauss}^2$, $\beta_1=27\pm2$Hz/Gauss and
$\beta_2=10 \pm2$Hz/Gauss. According to eq. (\ref{eqM}) this
amplitude of the oscillations corresponds to an imaginary conductance
of the order of $2.5\;10^{-3}\Omega^{-1}$ per ring for which the estimated
Drude conductance is $G_0=5.\;10^{-4}\Omega^{-1}$. This is clearly shown in
fig.\ref{fig_fQPhi}. The temperature dependence of the $h/2e$
periodic component of the signal (see fig. \ref{fig_fQT}) is
compatible with an exponential decay with a characteristic energy of
$200$mK over a range of temperature corresponding  to $1.5 \Delta
- 2E_c$.

\figtrois
 Since the frequency is smaller than the level spacing, we can assume that
the variations of $f_1$ reflect only the dc orbital magnetism of the
rings:
$-\frac{\partial f_1}{\partial H}\propto \frac{\partial^2
I_{per}}{\partial\Phi^2}$.
Thus, the amplitude of the oscillations
corresponds to a value of $1.5$nA per ring which is of the order of
$2E_c/\Phi_0=1.4$nA (the factor $2$ standing for spin), to be
compared to $2\sqrt{E_c\Delta}/\Phi_0=0.5$nA or
$2\Delta/\Phi_0=0.2$nA. However, under the same assumption,
our result implies a diamagnetic zero field persistent current.
One possible explanation of this sign could
be the presence of interactions which
modify the orbital magnetism of the rings
\cite{art_AmbEck}. However, the existence of an attractive interaction
between electrons in the $GaAs/GaAlAs$ heterojunction, which is
necessary to explain our experimental results, is not very likely,
but it cannot be completely ruled out. Furthermore we cannot exclude
that we are in the regime $\gamma<\omega$, implying a possible contribution
from relaxation processes
to $\Im m(G)$, but according to ref.\cite{art_RB}, this last
hypothesis does not explain the observed sign either.

	The magnetic field oscillations of the dissipative conductance,
obtained by integration of $\frac{\partial Q_1}{\partial H}$,
are presented in fig.\ref{fig_fQPhi}.
Their period is also $h/2e$, and their amplitude,
which is an order
of magnitude smaller than the oscillations of $\Im m(G)$,
corresponds to $0.2G_0$, i.e. 10 times
greater than the AAS $h/2e$ periodic weak localisation
magnetoconductance, as measured
in identical connected rings. This is the first indication that the
conductance measured in isolated rings is different from that
measured in connected ones.
The temperature dependence of the $h/2e$
periodic component of $\frac{\partial Q_1}{\partial H}$ depicted in
fig.\ref{fig_fQT} is notably different from the temperature
dependance of $\frac{\partial f_1}{\partial H}$, it is nearly independent of
temperature until $T=200$mK and strongly decreases at higher
temperature.
It is
not very easy to determine the sign of $\frac{\partial Q}{\partial H}$, since
we do not know very well the physical mechanisms at the origin of the
background signal which is dominated by the residual losses of the
$GaAs/GaAlAs$ substrate. We have nevertheless made a tentative
determination of this sign and found a negative magnetoconductance in
low field, which is opposite to the AAS result.
The observation of a negative sign for the low field
magnetoconductance at $T>\Delta$
could be another indication that $\gamma<\omega$\cite{art_RB}. In this limit,
interlevel absorption processes are the dominant contribution to $\Re e(G)$.
The magnetoconductance is then
related to the flux dependant spectral properties of the rings
and, according to random matrix theory, has negative sign in low field.
These results invite us to pursue the
experiment in a wider range of frequency and temperature.
\figquatre

Up to now we have only discussed the magnetic coupling of the rings
to the detector. However we also expect the resonator to be affected
by the polarisability of the rings, which experience a strong
electric field in their plane. This field is screened by the
electrons on a scale of the order of the Thomas-Fermi length.
However, since it is not infinitely small (of the order of 200\AA)
compared to the dimensions of the ring, one cannot exclude the
existence of quantum corrections to the polarisability of the rings
contributing to the flux dependence of $f_1$ with Aharonov-Bohm-like
oscillations. A quantitative estimation of this effect is clearly
needed and is underway.

In conclusion, we have designed an experiment sensitive enough to
measure the complex conductance of an array of rings in the relevant
frequency range. We have shown evidence of
$h/2e$ flux oscillations on both real and imaginary part of the conductance.
The
sign of the oscillations of the imaginary conductance corresponds to
diamagnetism in low field, which is for the moment difficult to
interprete. The periodic component of the real part of the
conductance has apparently a sign opposite to the weak localisation
oscillations
measured in connected rings, and is at least 10 times larger.

This work has strongly benefited from the help and suggestions of  L.
P. L\'evy, M. Nardonne, P. Pari and C. Urbina.
This work was partly supported by grant from DRET No. 92/181.

\end{document}